\documentclass[submission,copyright,creativecommons]{eptcs}
\providecommand{\event}{FESCA 2014} 

\usepackage{breakurl}             
\usepackage{amssymb}
\usepackage{graphicx}
\usepackage{amsmath,amssymb}
\usepackage{graphicx}
\usepackage{algorithm}
\usepackage{algorithmic}
\usepackage{multicol}
\usepackage{fancyvrb}
\usepackage[table]{xcolor}
\usepackage{esvect}
\usepackage{tikz}
\usepackage{stmaryrd}
\usepackage{setspace}
\title{Enabling Automatic Certification of Online Auctions}

\author{Wei Bai
\institute{Department of Computer Science \\
University of Liverpool \\
England, UK}
\email{Wei.Bai@liverpool.ac.uk}
\and
Emmanuel M. Tadjouddine
\institute{Department of Computer Science and Software Engineering \\
Xi'an Jiaotong-Liverpool University \\
SIP, Suzhou, China}
\email{Emmanuel.Tadjouddine@xjtlu.edu.cn}
\and
Yu Guo
\institute{School of Computer Science and Technology \\  University of Science and Technology of China \\ Hefei, China}
\email{guoyu@ustc.edu.cn}
}
\def\titlerunning{Enabling Automatic Certification of Online Auctions}
\def\authorrunning{W. Bai, E. M. Tadjouddine \& Y. Guo}

\newcommand{\coq}{\textsc{Coq}}

\newcommand{\owls}{\textsc{OWL-S}}
\newcommand{\owl}{\textsc{OWL-DL}}
\newcommand{\ruleML}{\textsc{RuleML}}
\newcommand{\mathML}{\textsc{MathML}}

\begin{document}
\maketitle

\begin{abstract}
We consider the problem of building up trust in a network of online auctions by software agents. This requires agents to have a deeper understanding of auction mechanisms and be able to verify desirable properties of a given mechanism. We have shown how these mechanisms can be formalised as semantic web services in \owls, a good enough expressive machine-readable formalism enabling software agents, to discover, invoke, and execute a web service. We have also used abstract interpretation to translate the auction's specifications from \owls\, based on description logic, to \coq, based on typed lambda calculus, in order to enable automatic verification of desirable properties of the auction by the software agents. For this language translation, we have discussed the syntactic transformation as well as the semantics connections between both concrete and abstract domains.  This work contributes to the implementation of the vision of agent-mediated e-commerce systems.
\end{abstract}

\section{Introduction}
\label{sec:intro}
Trust is an important issue in building up markets for rational participants. At the heart of markets lie the mechanism that spells out the rules of engagements for any participant. If the rules lack robustness, then paticipants may exploit the perceived loop holes to their benefit. To illustrate this statement, let us consider the  example of the LIBOR (London Inter Bank Offered Rate) scandal, see for example~\cite{mcandrews2008effect}. The LIBOR relies on a simple mechanism wherein leading banks in London report estimates of interest rates that they would be charged if they borrow from other banks; we then drop the four highest and the four lowest rates and then calculate the arithmetic mean of the remaining rates. This mechanism clearly relies upon trust: banks will report their rates truthfully. However, a small change in the LIBOR rate could generate large payments to a bank giving incentives to certain banks to influence the LIBOR rate. The so-called LIBOR scandal arose from a misreporting of interest rates by collusion and manipulation. The question is can we ensure that participants are truthful in their reports? In a much more general setting, can we ensure trust in the market?

In this paper, we consider electronic markets wherein software agents can buy or sell goods through online auction houses. This implies software agents will be engaging  financial transactions and entering legally binding contracts on behalf of their owners. We would like to enable agents to check that the auction house is trustworthy before entering it and bid for items on sale. For that purpose, we assume that electronic institutions publish the specifications of their  protocols in a high-level language specifying who can bid, in what order, and how the winners and payments are determined. A roaming agent who arrives at a foreign institution can download a protocol and analyze it in order to make a decision about whether or not to participate, and what strategy to use.

To illustrate these requirements, we need to bear in mind that an auction is typically defined by a winner determination algorithm; there are several in the literature, see~\cite{Cramton06PAT}. This provides the winning criteria and therefore implicitly determines the optimal strategy for participating agents. These winner determination algorithms are designed  with desirable properties such as truthful bidding is optimal.
If an electronic auction house is using such a protocol, it is not sufficient for the auctioneer to claim that a given bidding strategy is optimal. The fact is that sincerity cannot be assumed in open systems. These kinds of requirements are necessary to fully exploit the potential of e-commerce systems and to eventually achieve the vision of agent mediated e-commerce. In summary, the motivation for this work comes from the belief that verifying certain properties of a trading platform are useful to agents in open e-commerce environments only if: i) their protocols can be published in a machine-readable form; ii) their properties can be automatically verified by a software agent. Our focus is to address these requirements by providing a suitable specification language and associated proof procedures.

The Semantic Web not only enables greater access to content but also to service on the Web. \owls~\cite{martin2004owl} ontology is a language to describe Web services. It can be used  to describe the properties and capabilities of a Web service in an unambiguous and computer-interpretable form. As a result, we have formalised an online auction as a semantic web service by using \owls\ since this provides  machine-understandable specifications for software agents. To enable automatic verification procedures by agents, we present a translation framework that maps \owls\  into \coq\ specifications so that verification can be carried out as we have shown in our previous work~\cite{Bai2013APC} wherein we have implemented a verification procedure using the proof-carrying code paradigm from within \coq. The \owls-to-\coq\ language translation relies upon the abstract interpretation approach~\cite{Cousot2004} allowing us to reason within \coq\ and then deduce properties for \owls. The contributions of this ongoing work are two-fold: i) we have provided a machine readable formalism for online auctions by using \owls\ thus enabling software agents to automatically discover, invoke and execute an online auction; ii) we have shown how to map \owls\ auction into \coq\ specifications by using abstract interpretation so as to enable automatic verification by relying upon a well-established theorem prover.

\section{Background and Model}
\label{sec:model}
Generally speaking a {\em combinatorial auction}~\cite{Cramton06PAT} is composed of a set $I$ of $m$ items to be sold to $n$ potential buyers. A bid is formulated as a pair $(B(x),b(x))$ in which
$B(x) \subseteq I$ is a bundle of items and $b(x) \in R^{+}$ is the price offer for the items in $B$. The {\em combinatorial auction problem} (CAP) is to find a set $X_0 \subseteq X$ such that, for a given a set of $k$ bids, $ X = \{ (B(x_1), b(x_1)), (B(x_2), b(x_2)), \ldots, (B(x_k), b(x_k)) \}$, the quantity $\sum_{x \in X_0} b(x)$ is maximal  subject to the constraints expressed as
for all $x_i, x_j \in X_0: \; B(x_i) \cap B(x_j) = \emptyset$ meaning an item can be found in only one accepted bid. We assume {\em free disposal} meaning that items may remain unallocated at the end of the auction.


As shown in~\cite{Rothkopf1998CMC}, the CAP is NP-hard implying that approximation
algorithms are used to find a near-optimal solutions or restrictions are imposed in order to find tractable instances of the CAP~\cite{Tennenholtz2000STC} wherein polynomial time algorithms can be found for a restricted class of combinatorial auctions. A combinatorial auction can be {\em sub-additive} (for all bundles
$B_i, B_j \subseteq I$ such that $B_i \cap B_j = \emptyset$, the price offer for $B_i \cup B_j$
is less than or equals to the sum of the price offers for $B_i$ and $B_j$) or {\em super-additive}
(for all $B_i, B_j \subseteq I$ such that $B_i \cap B_j = \emptyset$, the price offer for $B_i \cup B_j$
is greater than or equals to the sum of the price offers for $B_i$ and $B_j$).

Game theory {\em mechanism}, see for example~\cite{Cramton06PAT}, is usually used to describe auctions, thus providing  decision procedures that determine the set of winners for the auction according to some desired objective.  An objective  may be that
the mechanism should maximise the {\em social welfare}, which can be for example the sum of all agents' utilities in the auction. Such a mechanism is termed {\em efficient}. For
open multi-agent systems, another desirable property for the mechanism designer can be
{\em strategyproofness} (truth telling is a dominant strategy for all agents). A mechanism that is strategyproof has a dominant strategy equilibrium.

A well known class of mechanisms that is efficient and  strategyproof  is the
Vickrey-Clarke-Groves (VCG), see for example~\cite{Cramton06PAT}.
The VCG mechanism is performed by finding (i) the allocation that maximises the social welfare
and (ii) a pricing rule allowing each winner to benefit from a discount according to his contribution
to the overall value for the auction. To formalise the VCG mechanism, let us introduce the following
notations:
\begin{itemize}
\item ${\cal X}$ is the set possible allocations
\item $v_i(x)$ is the true valuation of $x\in {\cal X}$ for bidder $i$
\item $b_i(x)$ is the bidding value of $x\in {\cal X}$ for bidder $i$
\item $x^*\in \mbox{argmax}_{x\in {\cal X}} \sum_{i=1}^{n} b_i(x)$ is the optimal allocation for the
submitted bids.
\item $x_{-i}^*\in \mbox{argmax}_{x\in {\cal X}} \sum_{j\neq i}^{n} b_j(x)$ is the optimal allocation if
agent $i$ were not to bid.
\item $u_i$ is the utility function for bidder $i$.
\end{itemize}
The VCG payment $p_i$ for bidder $i$ is defined as
\begin{equation}
\label{eq:vcgpaye}
\begin{array}{ll}
p_i & =  b_i(x^*) - \left (\sum_{j=1}^{n}b_j(x^*) - \sum_{j=1, \; j\neq i}^n b_j(x_{-i}^*) \right ) \\
    & =  \sum_{j=1, \; j\neq i}^n b_j(x_{-i}^*) - \sum_{j=1, \;j\neq i}^{n}b_j(x^*),
    \end{array}
\end{equation}
and then, the utility $u_i$ for agent $i$ is a quasi-linear function of
its valuation $v_i$ for the received bundle and payment $p_i$.
\begin{equation}
\label{eq:vcgu}
u_i(v_i, p_i) = v_i -p_i.
\end{equation}

Let $p_i^*$ and $p_i$ be the payments for agent $i$ when it bids its true valuation $v_i$ and any number $b_i$ respectively. Note $p_i, p_i^{*}$ are functions of $b_{-i}$.
The strategyproofness of the mechanism amounts to the following verification:
\begin{equation}
\label{eq:dse}
\forall i,\; \forall v_i, \; \forall b_i \;
u_i (v_i, p_i^*(b_{-i})) \ge u_i (v_i, p_i(b_{-i})).
\end{equation}
In the equation~(\ref{eq:vcgpaye}), the quantity $\sum_{j=1, \, j\neq i}^n b_j(x_{-i}^*)$ is
called the {\em Clark tax}. Another interesting property
of this VCG mechanism is that it is {\em weakly budget balanced}~\cite{Clarke1971}
meaning the sum of all payments is greater than or equal to zero.
For the VCG properties (e.g. strategyproof) to hold, the auctioneer must solve $n{+}1$ hard combinatorial
optimisation problems (the optimal allocation in the presence of all bidders followed by $n$ optimal allocations with each bidder removed) exactly.

{\em Single item auctions} are a particular case of combinatorial auctions wherein one item is sold at a time. Common single item auctions are the English, Dutch or Vickrey auctions, see for example~\cite{Cramton06PAT}. For the Vickrey auction, the payment for agent $i$ is defined as:
\begin{equation}
\label{eq:vpaye}
\begin{array}{ll}
p_i & =  b_i(x^*) - \left (\sum_{j=1}^{n}b_j(x^*) - \sum_{j=1, \; j\neq i}^n b_j(x_{-i}^*) \right ) \\
    & =  \sum_{j=1, \; j\neq i}^n b_j(x_{-i}^*) \; = \; \;\mbox{the second highest bid}.
    \end{array}
\end{equation}
For the English auction, the payment for agent $i$ is defined as:
\begin{equation}
\label{eq:epaye}
\begin{array}{ll}
p_i & =   b_i(x^*) \; = \; \mbox{the highest bid}.
    \end{array}
\end{equation}
The utilities of the Vickrey and English auctions are defined as in equation~(\ref{eq:vcgu}).

We assume that an online auction will run over a fixed time period $T$ and that the seller may have a  reserve revenue under which items cannot be sold. This reserve revenue can typically be the reserve price for single item auctions. After a bid, the seller waits for some time before accepting the bid if it yiels a revenue that is greater or equal to the reserve one or waits for the expiry time before deciding whether to reject or accept the bid. In this work, we aim at describing online auctions by using the semantic web, e.g., the \owls\ specification language and then translate the resulting description into \coq\ specifications in order to carry out the verification of desirable properties.

\section{\owls\ Specifications}
\label{sec:specs}
Online auctions of software agents is a good application wherein data can be processed by automated reasoning tools. Logic-based languages are useful tools to model and reason about systems. They allow us to specify behavioral requirements of components of a system and formulate desirable properties for an individual component or the entire system. The semantic web enables us to describe and reason about web services by using  {\em ontologies}. Ontologies permit us to describe the semantics of terms representing an area of knowledge. OWL, a W3C standard, is a description logic-based language that enables us to describe ontologies by using basic constructs such as concept definitions and relations between them. It has been used in a wide range of areas including biology, medicine, or aerospace~\cite{dadzie2009applying}. In this work,
we view an auction as a web service with enough logical attachments enabling a software agent to understand the auction rules and to carry out verifications of claimed properties.
Logic-based languages are usually chosen for their {\em expressivity} or on the fact that their underlying logic is {\em sound}, {\em complete} or {\em decidable}. Expressivity provides us with powerful constructs to describe things that may not be otherwise expressed. Soundness ensures that if a property $\phi$ can be deduced from a system (a set of statements) $\Gamma$ ($\Gamma \vdash \phi$), then $\phi$ is true as long as $\Gamma$ is ($\Gamma \models \phi$). Completeness states that any true statement can be established by proof steps in the logic's calculus. Formally $\Gamma \models \phi$ implies $\Gamma \vdash \phi$. A logic is decidable if for any statement we can construct an algorithm that decides if it is true or false.

\subsection{Auction Ontology}
Ontology can be used to formally describe the semantics of terms representing an area of knowledge and give explicit meaning to the information. This allows for automated reasoning, semantic search and knowledge management in a specific area of knowledge. The ontology of auction domain contains the following constructs: classes, relations, axioms, individuals and assertions. We have used the ontology language \owls, which is used to describe Semantic Web Services within the OWL-based framework in order to build up the auction ontology.

\begin{figure*}[htbp]
    \centering
    \begin{tabular}{c}
        \includegraphics[width=0.8\textwidth]{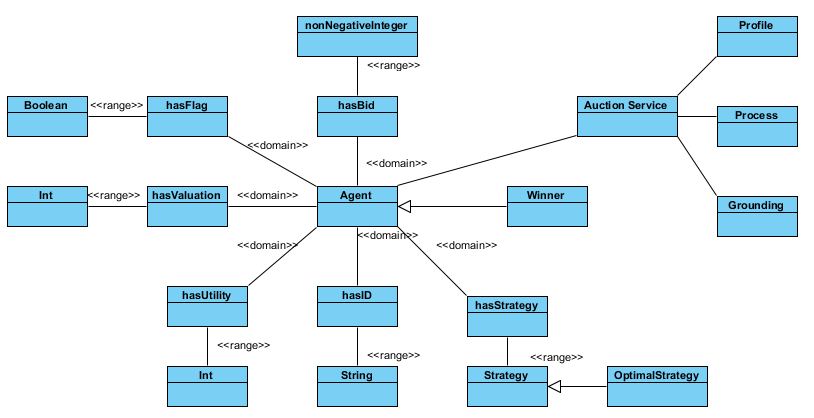}
    \end{tabular}
    \caption{Ontology for an auction service}
    \label{fig:topclass}
\end{figure*}
Figure~\ref{fig:topclass} displays the top level concepts of our online auction ontology.
The top-level classes are {\tt Agent} and {\tt AuctionService}. The class {\tt Winner} is a subclass of the class {\tt Agent}. The relations between classes are defined as properties. There are two kinds of properties: object properties and datatype properties. The object properties link individuals to individuals while the datatype properties link individuals to data values.
In the following section, we show how to encode this ontology within the \owls\ language.

\subsection{The \owls\ Description Language}
In this section, we argue that we can describe online auction houses as web services by using the \owls\ language, which provides us with a machine readable formalism and logical reasoning capabilities for software agents. We will start by showing that \owls\ is expressive enough to enable us to describe online auction mechanisms.
\newline
\begin{minipage}{0.1\textwidth}
\begin{tikzpicture}
\node(A) at (0,5){$$};
\node(B) at (0,0){$$};
\draw[->, >=latex, black!20!white, line width=5pt]   (A) to node[black,sloped,below]{expressiveness} (B) ;

\end{tikzpicture}
\end{minipage}%
\hfill
\begin{minipage}{0.95\textwidth}
\rowcolors{1}{black!5!white}{}

\begin{tabular}{cc}
   \parbox[t]{5cm}{\small XML: Syntactic, No Logic\\``Agent buyer has bid 100."} \\

\end{tabular}

\rowcolors{1}{black!15!white}{}

\begin{tabular}{cc}

  \parbox[t]{6cm}{\small OWL-DL: Description Logic\\sound, complete, decidable\\``Auction A has at leat 2 buyers."} \\
\end{tabular}

\rowcolors{1}{black!25!white}{}

\begin{tabular}{cc}

  \parbox[t]{7cm}{\small SWRL: OWL-DL + RuleML (including MathML and Horn rules)\\sound but not decidable\\``Utility = Valuation - Payment"} \\
\end{tabular}

\rowcolors{1}{black!35!white}{}

\begin{tabular}{cc}

  \parbox[t]{8cm}{\small OWL-S: SWRL + Programming Constructs\\sound but not decidable\\``While newbid \textgreater currentbid Do"} \\
\end{tabular}
\end{minipage}%

To describe online auctions, we may ask why not XML (Extensible Markup Language) as a description language for this task. XML provide a syntactic approach but no logical base for reasoning. The meaning of the relationships between XML elements can not be encoded. A language that builds upon XML and allows for reasoning is the \owl\ (Web Ontology Language), which is based on DL (Description Logic). Description logics are a family of logics that are decidable fragments of first-order logic. \owl\ is sound, complete, and decidable but with limited expressivity. For example, we cannot express the statement that 'an agent's utility is its valuation minus its payment'.
To extend \owl, the SWRL (Semantic Web Rule Language) combines \owl\ with \ruleML\ that includes among others, \mathML\ and Horn rules. As a result, SWRL is more expressive than \owl\ but SWRL is not decidable~\cite{horrocks2004swrl}. However, in SWRL, we cannot express the statement that 'while newbid \textgreater currentbid do'. Furthermore, auction mechanisms can be viewed as functions with inputs, outputs, preconditions or postconditions. They may contain complex programming constructs such as branching or iterations. \owls\ enables us to describe online auctions as Web Services.

An \owls\ description is mainly composed of a service \textit{profile} for advertising and discovering services; a \textit{process} model, which describes the operation of a service; and the \textit{grounding}, which specifies how to access a service. In our case, the process contains information about inputs, outputs, and a natural language description of the auction, e.g., this is a Vickrey auction. The grounding contains information on the service location so that an agent can run the service by using the \owls\ API. The process model is described as follows:\\

\begin{minipage}[h]{7cm}
{\scriptsize
\begin{verbatim}
define composite process Auction
(inputs: (...)
outputs: (...)
preconditions: (...)
results: (...)
)
{ // Process's Body
WinDetermAlgo(...);
PayeAndUtil(...) }
\end{verbatim}}
 \end{minipage}
 \begin{minipage}[h]{7cm}
{\scriptsize
\begin{verbatim}
<process:CompositeProcess rdf:ID="Auction">
<process:hasInput> ...
<process:hasOutput>
<process:hasPrecondition> ...
<process:hasResult> ...
...
<process:AtomicProcess rdf:ID="WinDetermAlgo"> ...
<process:AtomicProcess rdf:ID="PayeAndUtil"> ...
</process:CompositeProcess>
\end{verbatim}}
\end{minipage}
\\

The auction process model is basically composed of inputs, outputs, preconditions, results and a composition of two processes, which are the winner determination algorithm {\tt WinDetermAlgo} and {\tt PayeAndUtil} that calculates the payments as well as the utilities for the buyer agents. In \owls, we can also express auction properties such as the one defined by Equation~(\ref{eq:dse}) by using composite processes to define utility calculation as a kind of function and {\tt perform} operations. Note that \owls\ allows to express the forall operation in a first-order logic formula. We omit the \owls\ code for Equation~(\ref{eq:dse}) for clarity and lack of space.

\subsection{Implementation Issues}
We have implemented this agent-mediated semantic web service in JADE (Java Agent DEvelopment) framework~\cite{bellifemine2005jade} by using a Yellow Pages service. This permits seller agents to publish their services, so that buyer agents can find and exploit them. All the information of the buyer agents, such as bid and valuation of each agent, are translated into an OWL file, which is imported by  the  \owls\ service file containing the auction mechanism. We run this service using the \owls\ API to ensure the auction runs as expected. Eventually, the seller agent sends the result to each participant. The ACL language~\cite{bellifemine2005jade} is used for communicating messages between the agents.

\section{From \owls\ to \coq\ Specifications}
\label{sec:translation}
In Section~\ref{sec:specs}, we have shown how online auction mechanisms can be described using the \owls\ description language in order to have machine-understandable protocols by software agents.  On the other hand, we have illustrated that auction mechanisms can be specified within \coq\ so as to develop machine-checkable proofs of desirable mechanism properties. The verification approach relies upon  the proof-carrying code approach to enable the agents to gain confidence at the auction house by automatically verifying desirable properties. To bridge the gap between these two processes, we propose to transform \owls\ into \coq\ specifications by
\begin{enumerate}
\item devise an interpretation that maps the \owls\ semantics into \coq's so as verified properties by \coq\ can be deduced in the \owls\ domain.
\item translate \owls\ into a tailored subset of \coq\ specifications so that an \owls\ program or logical formula can be transformed into a \coq\ one in an automated fashion.
\end{enumerate}
To carry out the above two tasks, we rely upon abstract interpretation, see for example~\cite{Cousot2004}. The \owls\ specifications provide the following:
(i) a semantic domain ${\cal D}$ of states formed by vocabularies in terms of classes, subclasses, and properties; (ii) for each \owls\ program, an operational semantics $\llbracket \rrbracket \subseteq {\cal D} \times {\cal D}$ with some initial domain ${\cal D}^0$, (iii) for a given program $p$, the collecting semantics $\llbracket p \rrbracket$ is defined as the set of reachable states of ${\cal D}$ starting by an initial state in ${\cal D}^0$. An abstraction $\alpha: (2^{\cal D}, \subseteq, \cap, \cup) \rightarrow ({\cal D}^{\sharp}, \subseteq^{\sharp}, \cap^{\sharp}, \cup^{\sharp})$ wherein ${\cal D}^{\sharp}$ is an abstract domain with a lattice structure enables us to map concrete elements into abstract ones. The partial orders $\subseteq$ and $\subseteq^{\sharp}$ model the precision of the concrete and abstract elements in $2^{\cal D}$ and ${\cal D}^{\sharp}$ respectively. To define the abstract semantics in the abstract domain, we use a concretisation function $\gamma: ({\cal D}^{\sharp}, \subseteq^{\sharp}, \cap^{\sharp}, \cup^{\sharp}) \rightarrow (2^{\cal D}, \subseteq, \cap, \cup)$ such that:
\[ \forall p \; \alpha(p) \subseteq^{\sharp} p^{\sharp} \Rightarrow \llbracket p \rrbracket \subseteq \gamma(p^{\sharp}).\]

The abstraction $\alpha$ can be semantics-preserving if it does not use approximations in the abstract domain. If approximations are used then, conservative analyses are usually used. In this case, it is required that $\alpha$ be at least a {\em sound} abstraction so as to ensure that for a given program and a property $f$, $ f^{\sharp} = \alpha (f) = \; \mbox{true} \; \Rightarrow f = \; \mbox{true}$.
If $f^{\sharp}$ is false in the abstract domain, then we pick the counter-example found in the abstract domain, translate it into the concrete domain and simulate it to discover if it is indeed a counter-example of the concrete program. If this is the case, then $f$ does not hold, see~\cite{gulavani2006counterexample}.

\subsection{Translating Specifications}
We have used classical set operations to define OWL descriptions such as classes, properties, individuals and data values within \coq. The following table  lists part of the mapping; more details are available upon request.

\begin{center}
    \begin{tabular}{| l | p{10cm} |}
    \hline

    OWL descriptions &
     \coq\ equivalent definitions  \\ \hline
    Individual &
    \scriptsize{\begin{minipage}{3in}
     \begin{verbatim}
     Variable Individual : Set.
     \end{verbatim}
        \end{minipage}}\\ \hline

    Class &
    \scriptsize{\begin{minipage}{3in}
     \begin{verbatim}
     Definition Class := set Individual.
     \end{verbatim}
        \end{minipage}} \\ \hline

    ObjectProperty &
    \scriptsize{\begin{minipage}{3in}
     \begin{verbatim}
     Definition ObjectProperty :=
        set (prod Individual Individual).
     \end{verbatim}
        \end{minipage}} \\ \hline

    DatatypeProperty &
    \scriptsize{ \begin{minipage}{3in}
     \begin{verbatim}
     Definition DataProperty :=
        set (prod Individual Value).
     \end{verbatim}
     \end{minipage}} \\ \hline

    SomeValuesFrom() &
    \scriptsize{ \begin{minipage}{3in}
     \begin{verbatim}
     Definition someValuesFrom (op: ObjectProperty)
      (c: Class) : Class :=
        fun i =>
        exists y, set_in (i, y) op /\ (y <- c).
     \end{verbatim}
        \end{minipage}} \\ \hline

    AllValuesFrom() &
    \scriptsize{ \begin{minipage}{3in}
     \begin{verbatim}
     Definition allValuesFrom (op: ObjectProperty)
      (c: Class) : Class :=
        fun i =>
        forall y, set_in (i, y) op -> (y <- c).
     \end{verbatim}
     \end{minipage}} \\ \hline

    \end{tabular}
\end{center}

By mapping the \owls\ constructs into \coq\ equivalent definitions, we can transform an \owls\ specification into a \coq\ counterpart as illustrated below.

\begin{minipage}[h]{7cm}
{\scriptsize
     \begin{verbatim}
     <Class IRI="#Agent"/>
        <Class IRI="#Agent"/>
        <NamedIndividual IRI="#agent_1"/>
        <Class IRI="#Agent"/>
        <NamedIndividual IRI="#agent_2"/>
     \end{verbatim}}
        \end{minipage}
    \begin{minipage}[h]{7cm}
{\scriptsize
\begin{verbatim}
Inductive Individual : Set :=
  | agent (n: nat).
Definition Agent : Class :=
  set_add (agent _2)
     (set_add (agent _1)empty_class).
\end{verbatim}}
\end{minipage} \\

In the \owls\ code on the left, we define a class {\tt Agent} and two individuals {\tt Agent\_1} and {\tt Agent\_2} that belongs to the class {\tt Agent}. On the right, we show what will be the result of translating the \owls\ code fragment into \coq. Such a translation will rely on compiler technology.

\subsection{Verifying Desirable Properties}
\label{sec:verif}
In previous work~\cite{Bai2013APC}, we have relied upon the Proof-Carrying Code (PCC) ideas since it allows us to shift the burden of proof from the buyer agent to the auctioneer who can spend time to prove a claimed property once for all so that it can be checked by a software agent willing to join the auction house. The certification procedure works as follows. The buyer agent arriving at the auction house can download its specification and the claimed proof of a desirable property. Then, the buyer installs the proof checker, which is a standalone verifier for \coq\ proofs. After the proof checker is installed to the consumer side, the buyer can now perform all verifications of claimed properties of the auction before deciding whether to join and with which bidding strategy.

\coq~\cite{TheCoqDevelopmentTeam2012} is an interactive theorem prover based on the calculus of inductive constructions allowing definitions of data types, predicates, and functions. It also enables us to express mathematical theorems and  to interactively develop proofs that are checked from within the system. We have used \coq\ because it has been developed for more than twenty years and is widely used for formal proof development in a variety of contexts related to software and hardware correctness and safety. For example,  \coq\ was used to develop and certify a compiler~\cite{Leroy2009}. A fully computer-checked proof of the Four Colour Theorem was created in~\cite{Gonthier2008}. In~\cite{Vestergaard2006}, a \coq-formalised proof that all non-cooperative and sequential games have a Nash equilibrium point is presented.
It turns out that \coq\ is a good example of combination of logic and computation as it allows for the formalisation of different types of logic (e.g., first order logic, etc.) while providing the possibility of defining functions, which are the cornerstone of computation.

A well-defined auction mechanism can be viewed as a function that maps a set of typed agents into outcomes characterised by utilities usually defined as pseudo-linear equations, see Section~\ref{sec:model}. Not only, we need to specify rules and properties but we also need to carry out some calculations. The constructive approach provided by \coq\ offers possibilities to describe auctions along with desirable properties and prove them. In~\cite{Bai2013APC}, we have developed the \coq\ proof of the well-known statement that bidding its true valuation is the dominant strategy for each agent in a Vickrey or English auction. Another important property that can be checked is that the auction is a well-defined function and that it does implement its specifications through certified code generation~\cite{Caminati2013proving}. More challenging properties to be checked might be that the auction mechanism is collusion-free or that it is free from fictitious bidding. In this work, we focus on the mapping of auction specifications from \owls\ to \coq\ so as to enable the verification in a more dedicated and powerful proof development tool.

\section{Related Work}\label{sec:rw}
There are several related works about auction ontologies. In~\cite{zou2003using}, a framework called TAGA is used to simulate the automated trading in dynamic markets.  TAGA uses semantic web language to specify and publish underlying common ontologies.  However, the specification of auction services in this paper does not describe the explicit process and rules of different auctions.  In~\cite{tamma2002ontology}, a  shared negotiation ontology is used to help agents negotiate with each other with a possible application to simulate online auction competition. The simulation was not a real implementation but just an illustration. In~\cite{subercaze2011programming}, the authors integrate semantic web technologies into agent architecture to represent the knowledge and behavior of agent. Although, this paper is mainly about distributed knowledge management, the presented approach can help us building up an agent mediated and automatically processed online auction service.

The idea of software agents automatically checking desirable properties has been investigated in~\cite{Tadjouddine08AfM} by using a model checking approach. The computational complexity of such costly verification procedures are investigated in~\cite{Tadjouddine11CCo}. More recently, a proof-carrying code approach relying on proof development tools to enable automatic certification of auction mechanisms has been investigated in~\cite{Bai2013APC,Caminati2013proving}. Our work is aimed at bridging the gap between the need for a machine-readable formalism to specify online auction mechanisms and the ability of software agents to check possibly complex auction properties.

In the context of specifications translation, the work reported in~\cite{lucanu2005institution} has illustrated a case of transformation from OWL to Z specifications. The soundness of this mapping is shown using an algebraic institution morphism by viewing the underlying logics of OWL and Z as institutions. In~\cite{miao2009formal}, OWL-S was mapped into Frame logic for the use of first order logic based model checking to verify certain properties of semantic web service systems. In~\cite{zaremski1997specification}, a matching approach was used to select appropriate component software in a computer program translation scheme by inspecting the component software specifications. In our work, we define a syntactic translation scheme followed by a semantics interpretation based on the abstract interpretation paradigm.
\section{Concluding Remarks}\label{sec:con}
In this paper, we have considered the problem of trust in a network of online auctions by software agents. We have focused on the specifications of auction mechanisms required to be machine-understandable and have proposed an \owls\ formalisation of the mechanisms. We have discussed how \owls\ is expressive enough for the description of these mechanisms. To enable automatic verification of auction properties by software agents, we have relied upon the well-established proof development \coq\ based on lambda calculus by translating the auction specifications from \owls\ to \coq. For this translation, we have illustrated the syntactic transformation and have used abstract interpretation to connect the concrete semantics of the \owls\ domain into the semantics of \coq.

Future work includes the followings. We first need to formally establish the soundness (possibly the completeness) of the translation of auction specifications from \owls\ to \coq. We will then build up the infrastructure that connects the agents simulation tool JADE to \coq\ in an automated fashion. In JADE, we should be able to simulate two or three auction services and a number of software agents capable of understanding those auctions thanks to their \owls\ description, translate their specifications from \owls\ to \coq\ so that proofs of desirable properties of those mechanisms can be developed from within \coq, and then finally, enable a software agent to connect to \coq\ in order to certify a given property by using the proof-carrying code as shown in~\cite{Bai2013APC}. Note that the translation from \owls\ to \coq\ relies upon compiler construction technology. Such an infrastructure will enable us to simulate {\em artificial markets} (since they are mediated by software agents) that will shed some light on real markets by making assumptions and analysing their effects on the specified artificial societies.


\begin{thebibliography}{10}
\providecommand{\bibitemdeclare}[2]{}
\providecommand{\surnamestart}{}
\providecommand{\surnameend}{}
\providecommand{\urlprefix}{Available at }
\providecommand{\url}[1]{\texttt{#1}}
\providecommand{\href}[2]{\texttt{#2}}
\providecommand{\urlalt}[2]{\href{#1}{#2}}
\providecommand{\doi}[1]{doi:\urlalt{http://dx.doi.org/#1}{#1}}
\providecommand{\bibinfo}[2]{#2}
\begin{spacing}{0.9}
\bibitemdeclare{conference}{Bai2013APC}
\bibitem{Bai2013APC}
\bibinfo{author}{Wei \surnamestart Bai\surnameend},
  \bibinfo{author}{Emmanuel~M. \surnamestart Tadjouddine\surnameend},
  \bibinfo{author}{Terry \surnamestart Payne\surnameend} \&
  \bibinfo{author}{Steven \surnamestart Guan\surnameend}
  (\bibinfo{year}{2013}): \emph{\bibinfo{title}{A Proof-Carrying Code Approach
  to Certificate Auction Mechanisms}}.
\newblock In: {\sl \bibinfo{booktitle}{The 10th International Symposium on
  Formal Aspects of Component Software}}.

\bibitemdeclare{incollection}{bellifemine2005jade}
\bibitem{bellifemine2005jade}
\bibinfo{author}{Fabio \surnamestart Bellifemine\surnameend},
  \bibinfo{author}{Federico \surnamestart Bergenti\surnameend},
  \bibinfo{author}{Giovanni \surnamestart Caire\surnameend} \&
  \bibinfo{author}{Agostino \surnamestart Poggi\surnameend}
  (\bibinfo{year}{2005}): \emph{\bibinfo{title}{JADE -- A Java Agent
  DEvelopment Framework}}.
\newblock In: {\sl \bibinfo{booktitle}{Multi-Agent Programming}},
  \bibinfo{publisher}{Springer}, pp. \bibinfo{pages}{125--147},
  \doi{10.1007/0-387-26350-0\_5}

\bibitemdeclare{article}{Caminati2013proving}
\bibitem{Caminati2013proving}
\bibinfo{author}{Marco~B \surnamestart Caminati\surnameend},
  \bibinfo{author}{Manfred \surnamestart Kerber\surnameend},
  \bibinfo{author}{Christoph \surnamestart Lange\surnameend} \&
  \bibinfo{author}{Colin \surnamestart Rowat\surnameend}
  (\bibinfo{year}{2013}): \emph{\bibinfo{title}{Proving soundness of
  combinatorial Vickrey auctions and generating verified executable code}}.
\newblock {\sl \bibinfo{journal}{arXiv preprint arXiv:1308.1779}}.

\bibitemdeclare{article}{Clarke1971}
\bibitem{Clarke1971}
\bibinfo{author}{Edward~H \surnamestart Clarke\surnameend}
  (\bibinfo{year}{1971}): \emph{\bibinfo{title}{Multipart pricing of public
  goods}}.
\newblock {\sl \bibinfo{journal}{Public choice}}
  \bibinfo{volume}{11}(\bibinfo{number}{1}), pp. \bibinfo{pages}{17--33},
  \doi{10.1007/BF01726210}.

\bibitemdeclare{article}{Cousot2004}
\bibitem{Cousot2004}
\bibinfo{author}{P.~\surnamestart Cousot\surnameend} \&
  \bibinfo{author}{R.~\surnamestart Cousot\surnameend} (\bibinfo{year}{2004}):
  \emph{\bibinfo{title}{Basic concepts of abstract interpretation}}.
\newblock {\sl \bibinfo{journal}{Building the Information Society}}, pp.
  \bibinfo{pages}{359--366},
  \doi{10.1007/978-1-4020-8157-6\_27}.

\bibitemdeclare{article}{Cramton06PAT}
\bibitem{Cramton06PAT}
\bibinfo{author}{Peter \surnamestart Cramton\surnameend}, \bibinfo{author}{Yoav
  \surnamestart Shoham\surnameend} \& \bibinfo{author}{Richard \surnamestart
  Steinberg\surnameend} (\bibinfo{year}{2006}):
  \emph{\bibinfo{title}{Combinatorial auctions}},ISBN:\bibinfo{isbn}{9780262033428}.

\bibitemdeclare{article}{dadzie2009applying}
\bibitem{dadzie2009applying}
\bibinfo{author}{A-S \surnamestart Dadzie\surnameend},
  \bibinfo{author}{R~\surnamestart Bhagdev\surnameend},
  \bibinfo{author}{A~\surnamestart Chakravarthy\surnameend},
  \bibinfo{author}{S~\surnamestart Chapman\surnameend},
  \bibinfo{author}{J~\surnamestart Iria\surnameend},
  \bibinfo{author}{V~\surnamestart Lanfranchi\surnameend},
  \bibinfo{author}{J~\surnamestart Magalh{\~a}es\surnameend},
  \bibinfo{author}{D~\surnamestart Petrelli\surnameend} \&
  \bibinfo{author}{F~\surnamestart Ciravegna\surnameend}
  (\bibinfo{year}{2009}): \emph{\bibinfo{title}{Applying semantic web
  technologies to knowledge sharing in aerospace engineering}}.
\newblock {\sl \bibinfo{journal}{Journal of Intelligent Manufacturing}}
  \bibinfo{volume}{20}(\bibinfo{number}{5}), pp. \bibinfo{pages}{611--623},
  \doi{10.1007/s10845-008-0141-1}.

\bibitemdeclare{article}{Gonthier2008}
\bibitem{Gonthier2008}
\bibinfo{author}{G.~\surnamestart Gonthier\surnameend} (\bibinfo{year}{2008}):
  \emph{\bibinfo{title}{The four colour theorem: Engineering of a formal
  proof}}.
\newblock {\sl \bibinfo{journal}{Computer Mathematics}}, pp.
  \bibinfo{pages}{333--333}, \doi{10.1007/978-3-540-87827-8\_28}.

\bibitemdeclare{incollection}{gulavani2006counterexample}
\bibitem{gulavani2006counterexample}
\bibinfo{author}{Bhargav~S \surnamestart Gulavani\surnameend} \&
  \bibinfo{author}{Sriram~K \surnamestart Rajamani\surnameend}
  (\bibinfo{year}{2006}): \emph{\bibinfo{title}{Counterexample driven
  refinement for abstract interpretation}}.
\newblock In: {\sl \bibinfo{booktitle}{Tools and Algorithms for the
  Construction and Analysis of Systems}}, \bibinfo{publisher}{Springer}, pp.
  \bibinfo{pages}{474--488}, \doi{10.1007/11691372\_34}.

\bibitemdeclare{article}{horrocks2004swrl}
\bibitem{horrocks2004swrl}
\bibinfo{author}{Ian \surnamestart Horrocks\surnameend},
  \bibinfo{author}{Peter~F \surnamestart Patel-Schneider\surnameend},
  \bibinfo{author}{Harold \surnamestart Boley\surnameend},
  \bibinfo{author}{Said \surnamestart Tabet\surnameend},
  \bibinfo{author}{Benjamin \surnamestart Grosof\surnameend},
  \bibinfo{author}{Mike \surnamestart Dean\surnameend} et~al.
  (\bibinfo{year}{2004}): \emph{\bibinfo{title}{SWRL: A semantic web rule
  language combining OWL and RuleML}}.
\newblock {\sl \bibinfo{journal}{W3C Member submission}} \bibinfo{volume}{21},{79}.


\bibitemdeclare{article}{Leroy2009}
\bibitem{Leroy2009}
\bibinfo{author}{X.~\surnamestart Leroy\surnameend} (\bibinfo{year}{2009}):
  \emph{\bibinfo{title}{Formal verification of a realistic compiler}}.
\newblock {\sl \bibinfo{journal}{Communications of the ACM}}
  \bibinfo{volume}{52}(\bibinfo{number}{7}), pp. \bibinfo{pages}{107--115},
  \doi{10.1145/1538788.1538814}.

\bibitemdeclare{inproceedings}{lucanu2005institution}
\bibitem{lucanu2005institution}
\bibinfo{author}{Dorel \surnamestart Lucanu\surnameend},
  \bibinfo{author}{Yuan-Fang \surnamestart Li\surnameend} \&
  \bibinfo{author}{Jin~Song \surnamestart Dong\surnameend}
  (\bibinfo{year}{2005}): \emph{\bibinfo{title}{Institution Morphisms for
  Relating OWL and Z.}}
\newblock In: {\sl \bibinfo{booktitle}{SEKE}}, pp. \bibinfo{pages}{286--291}.

\bibitemdeclare{article}{martin2004owl}
\bibitem{martin2004owl}
\bibinfo{author}{David \surnamestart Martin\surnameend}, \bibinfo{author}{Mark
  \surnamestart Burstein\surnameend}, \bibinfo{author}{Jerry \surnamestart
  Hobbs\surnameend}, \bibinfo{author}{Ora \surnamestart Lassila\surnameend},
  \bibinfo{author}{Drew \surnamestart McDermott\surnameend},
  \bibinfo{author}{Sheila \surnamestart McIlraith\surnameend},
  \bibinfo{author}{Srini \surnamestart Narayanan\surnameend},
  \bibinfo{author}{Massimo \surnamestart Paolucci\surnameend},
  \bibinfo{author}{Bijan \surnamestart Parsia\surnameend},
  \bibinfo{author}{Terry \surnamestart Payne\surnameend} et~al.
  (\bibinfo{year}{2004}): \emph{\bibinfo{title}{OWL-S: Semantic markup for web
  services}}.
\newblock {\sl \bibinfo{journal}{W3C member submission}} \bibinfo{volume}{22},{2007--04}.


\bibitemdeclare{techreport}{mcandrews2008effect}
\bibitem{mcandrews2008effect}
\bibinfo{author}{James \surnamestart McAndrews\surnameend},
  \bibinfo{author}{Asani \surnamestart Sarkar\surnameend} \&
  \bibinfo{author}{Zhenyu \surnamestart Wang\surnameend}
  (\bibinfo{year}{2008}): \emph{\bibinfo{title}{The effect of the term auction
  facility on the london inter-bank offered rate}}.
\newblock \bibinfo{type}{Technical Report}, \bibinfo{institution}{Staff Report,
  Federal Reserve Bank of New York}, \doi{10.2139/ssrn.1183671}.

\bibitemdeclare{incollection}{miao2009formal}
\bibitem{miao2009formal}
\bibinfo{author}{Huaikou \surnamestart Miao\surnameend}, \bibinfo{author}{Tao
  \surnamestart He\surnameend} \& \bibinfo{author}{Liping \surnamestart
  Li\surnameend} (\bibinfo{year}{2009}): \emph{\bibinfo{title}{Formal semantics
  of OWL-S with F-logic}}.
\newblock In: {\sl \bibinfo{booktitle}{Computer and Information Science 2009}},
  \bibinfo{publisher}{Springer}, pp. \bibinfo{pages}{105--117},
  \doi{10.1007/978-3-642-01209-9\_10}.

\bibitemdeclare{article}{Rothkopf1998CMC}
\bibitem{Rothkopf1998CMC}
\bibinfo{author}{Michael~H \surnamestart Rothkopf\surnameend},
  \bibinfo{author}{Aleksandar \surnamestart Peke{\v{c}}\surnameend} \&
  \bibinfo{author}{Ronald~M \surnamestart Harstad\surnameend}
  (\bibinfo{year}{1998}): \emph{\bibinfo{title}{Computationally manageable
  combinational auctions}}.
\newblock {\sl \bibinfo{journal}{Management science}}
  \bibinfo{volume}{44}(\bibinfo{number}{8}), pp. \bibinfo{pages}{1131--1147},
  \doi{10.1287/mnsc.44.8.1131}.

\bibitemdeclare{incollection}{subercaze2011programming}
\bibitem{subercaze2011programming}
\bibinfo{author}{Julien \surnamestart Subercaze\surnameend} \&
  \bibinfo{author}{Pierre \surnamestart Maret\surnameend}
  (\bibinfo{year}{2011}): \emph{\bibinfo{title}{Programming Semantic Agent for
  Distributed Knowledge Management}}.
\newblock In: {\sl \bibinfo{booktitle}{Semantic Agent Systems}},
  \bibinfo{publisher}{Springer}, pp. \bibinfo{pages}{47--65},
  \doi{10.1007/978-3-642-18308-9\_3}.

\bibitemdeclare{article}{Tadjouddine11CCo}
\bibitem{Tadjouddine11CCo}
\bibinfo{author}{Emmanuel~M. \surnamestart Tadjouddine\surnameend}
  (\bibinfo{year}{2011}): \emph{\bibinfo{title}{Computational Complexity of
  Some Intelligent Computing Systems}}.
\newblock {\sl \bibinfo{journal}{International Journal of Intelligent Computing
  and Cyberneticsl}} \bibinfo{volume}{4}(\bibinfo{number}{2}), pp.
  \bibinfo{pages}{144 -- 159}, \doi{10.1108/17563781111136676}.

\bibitemdeclare{incollection}{Tadjouddine08AfM}
\bibitem{Tadjouddine08AfM}
\bibinfo{author}{Emmanuel~M \surnamestart Tadjouddine\surnameend},
  \bibinfo{author}{Frank \surnamestart Guerin\surnameend} \&
  \bibinfo{author}{Wamberto \surnamestart Vasconcelos\surnameend}
  (\bibinfo{year}{2009}): \emph{\bibinfo{title}{Abstracting and Verifying
  Strategy-Proofness for Auction Mechanisms}}.
\newblock In: {\sl \bibinfo{booktitle}{Declarative Agent Languages and
  Technologies VI}}, \bibinfo{publisher}{Springer}, pp.
  \bibinfo{pages}{197--214}, \doi{10.1007/978-3-540-93920-7\_13}.

\bibitemdeclare{incollection}{tamma2002ontology}
\bibitem{tamma2002ontology}
\bibinfo{author}{Valentina \surnamestart Tamma\surnameend},
  \bibinfo{author}{Michael \surnamestart Wooldridge\surnameend},
  \bibinfo{author}{Ian \surnamestart Blacoe\surnameend} \& \bibinfo{author}{Ian
  \surnamestart Dickinson\surnameend} (\bibinfo{year}{2002}):
  \emph{\bibinfo{title}{An ontology based approach to automated negotiation}}.
\newblock In: {\sl \bibinfo{booktitle}{Agent-Mediated Electronic Commerce IV.
  Designing Mechanisms and Systems}}, \bibinfo{publisher}{Springer}, pp.
  \bibinfo{pages}{219--237}, \doi{10.1007/3-540-36378-5\_14}.

\bibitemdeclare{inproceedings}{Tennenholtz2000STC}
\bibitem{Tennenholtz2000STC}
\bibinfo{author}{Moshe \surnamestart Tennenholtz\surnameend}
  (\bibinfo{year}{2000}): \emph{\bibinfo{title}{Some tractable combinatorial
  auctions}}.
\newblock In: {\sl \bibinfo{booktitle}{AAAI/IAAI}}, pp.
  \bibinfo{pages}{98--103}.

\bibitemdeclare{misc}{TheCoqDevelopmentTeam2012}
\bibitem{TheCoqDevelopmentTeam2012}
\bibinfo{author}{\surnamestart {The Coq Development Team}\surnameend}
  (\bibinfo{year}{2012}): \emph{\bibinfo{title}{The Coq proof assistant
  reference manual: Version 8.4}}.
\newblock \bibinfo{howpublished}{\url{http://coq.inria.fr}}.

\bibitemdeclare{article}{Vestergaard2006}
\bibitem{Vestergaard2006}
\bibinfo{author}{R.~\surnamestart Vestergaard\surnameend}
  (\bibinfo{year}{2006}): \emph{\bibinfo{title}{A constructive approach to
  sequential Nash equilibria}}.
\newblock {\sl \bibinfo{journal}{Information Processing Letters}}
  \bibinfo{volume}{97}(\bibinfo{number}{2}), pp. \bibinfo{pages}{46--51},
  \doi{10.1016/j.ipl.2005.09.010}.

\bibitemdeclare{article}{zaremski1997specification}
\bibitem{zaremski1997specification}
\bibinfo{author}{Amy~Moormann \surnamestart Zaremski\surnameend} \&
  \bibinfo{author}{Jeannette~M \surnamestart Wing\surnameend}
  (\bibinfo{year}{1997}): \emph{\bibinfo{title}{Specification matching of
  software components}}.
\newblock {\sl \bibinfo{journal}{ACM Transactions on Software Engineering and
  Methodology (TOSEM)}} \bibinfo{volume}{6}(\bibinfo{number}{4}), pp.
  \bibinfo{pages}{333--369}, \doi{10.1145/261640.261641}.

\bibitemdeclare{inproceedings}{zou2003using}
\bibitem{zou2003using}
\bibinfo{author}{Youyong \surnamestart Zou\surnameend}, \bibinfo{author}{Tim
  \surnamestart Finin\surnameend}, \bibinfo{author}{Li~\surnamestart
  Ding\surnameend}, \bibinfo{author}{Harry \surnamestart Chen\surnameend} \&
  \bibinfo{author}{Rong \surnamestart Pan\surnameend} (\bibinfo{year}{2003}):
  \emph{\bibinfo{title}{Using semantic web technology in multi-agent systems: a
  case study in the TAGA trading agent environment}}.
\newblock In: {\sl \bibinfo{booktitle}{Proceedings of the 5th international
  conference on Electronic commerce}}, \bibinfo{organization}{ACM}, pp.
  \bibinfo{pages}{95--101}, \doi{10.1145/948005.948018}.
\end{spacing}
\end{thebibliography}

\end{document}